\documentclass[showpacs,preprint,pra]{revtex4}
\usepackage{dcolumn,longtable}
\def\rhoo{\mbox{\boldmath{$\rho$}}}
\begin{document}
\title{
Accurate long-range coefficients for two excited like isotope He
atoms: He($2\,^1\!P$)--He($2\,^1\!P$),
He($2\,^1\!P$)--He($2\,^3\!P$), and He($2\,^3\!P$)--He($2\,^3\!P$)
}
\author{J.-Y. Zhang,\footnote{Present address: Faculty of Technology,
 Charles Darwin University, Northern Territory, Australia 0909}$^1$ Z.-C. Yan,$^{1, \, \, 2}$ D. Vrinceanu,$^{3}$ J. F. Babb,$^4$ and H. R.
Sadeghpour$^4$ }
\affiliation{$^1$Department of Physics,
University of New Brunswick, Fredericton, New Brunswick, Canada
E3B 5A3}
\affiliation{$^2$Shanghai United Center for Astrophysics, Shanghai
Normal University,
100 Guilin Road, Shanghai, People's Republic of China 200234}
\affiliation{$^3$Theoretical Division, Los Alamos National
Laboratory, Los Alamos, New Mexico 87545, USA}
\affiliation{$^4$ITAMP, Harvard-Smithsonian Center for
Astrophysics, Cambridge, Massachusetts 02138, USA}
\date{\today}

\begin{abstract}
A general formalism is used to express the long-range potential
energies in inverse powers of the separation distance between two
like atomic or molecular systems with $P$ symmetries.  The
long-range molecular interaction coefficients are calculated for
the molecular symmetries $\Delta$, $\Pi$, and $\Sigma$,  arising
from the following interactions: He($2\,^1\!P$)--He($2\,^1\!P$),
He($2\,^1\!P$)--He($2\,^3\!P$), and
He($2\,^3\!P$)--He($2\,^3\!P$). The electric quadrupole-quadrupole
term, $C_{5}$, the van der Waals (dispersion) term $C_{6}$, and
higher-order terms, $C_{8}$, and $C_{10}$, are calculated
\textit{ab initio} using accurate variational wave functions in
Hylleraas coordinates with finite nuclear mass effects. A
comparison is made with previously published results where
available.
\end{abstract}

\pacs{32.10.Dk,34.20.Cf} \maketitle

\section {Introduction}
Accurate description of the interactions between two excited atoms
(or molecules) at long-range is  fundamentally important for
studies of molecular excited state
spectroscopy~\cite{DulLevMag96,BoiSimCot02}, associative
ionization~\cite{HeaJul93,AmeJonLet00}, and other collisional
processes~\cite{DerPorKot03}, and is at the heart of several
schemes for quantum computation~\cite{Singer05,WalSaf05}.  At
sufficiently large separations, the mutual electrostatic
interaction energy between the two excited atoms can be accurately
described using an expansion of the potential energy in inverse
powers of the separation distance $R$.  The terms describe the
electric quadrupole-quadrupole interaction at order $R^{-5}$ and
the instantaneous dipole-dipole (e.g.  dispersion) interaction at
order $R^{-6}$~\cite{chang67} and higher order instantaneous
multipole-multipole interactions at orders $R^{-8}$ and $R^{-10}$.

Long-range interactions involving few-electron atoms are the only
interactions that
presently can be rigorously calculated with high accuracy.
Different levels of approximation are needed for the calculations
of long-range forces for alkali-metal and alkaline-earth
atoms.~\cite{Ovs82,marinescu97,DerPorKot03}. Sizeable
discrepancies between various calculations in the literature can
occur, as illustrated in the comparisons of $C_6$ coefficients,
for example, given by Zhang \textit{ et
  al.}~\cite{ZhaMitBro07} for Li $(2p)$--Li $(2p)$ and given by
Yurova~\cite{Yur02} for Na$(3p)$--Na$(3p)$. For helium, it is
possible to  perform a highly-accurate \textit{ ab initio}
calculation of atomic properties and long-range interaction
coefficients.  Such results could become benchmarks for eventual
{\it ab initio} calculations of alkaline-earth atomic
interactions. Alkaline-earth and other two-electron excited P
atoms are currently being studied as the optimal candidates for
frequency-based standards and optical clock experiments
\cite{ITAMPWorkshop06}.

We had previously studied the long-range interaction coefficients
$C_n$ (with $n\leq 10$) for
 all He($n\,^\lambda S$)--He($n'\,^{\lambda'} S$) and He($n\,^\lambda S$)--He($n'\,^{\lambda'} P$) systems of the energetically lowest
five states: He($1^1S$), He($2^3S$), He($2^1S$), He($2^3P$) and
He($2^1P$) and the finite nuclear mass effects for like
isotopes~\cite{sadeg2,jzd02,jzd03}. In this work, we present
results for more complicated set of interactions between two like
isotope helium atoms with $P$ symmetries. Degenerate perturbation
theory is needed to derive the interaction terms for some of the
terms. Section \textbf{II} introduces a general formalism for
calculating dispersion coefficients between two like atomic or
molecular systems of $P$ symmetry. Section \textbf{III} presents
numerical results of dispersion coefficients $C_5$, $C_6$, $C_8$,
and $C_{10}$ for the following three systems
He$(2\,^1\!P)$--He$(2\,^1\!P)$,
 He$(2\,^1\!P)$--He$(2\,^3\!P)$,
and He$(2\,^3\!P)$--He$(2\,^3\!P)$.
\section {Formulation}
In this work,  atomic units are used throughout. At large
distances $R$ between two atoms $a$ and $b$, the Coulomb
interaction~\cite{yan96}, treated as a perturbation to the two
isolated atoms, is
\begin{eqnarray}
V &=& \sum_{\ell=0}^{\infty}\sum_{L=0}^{\infty} \frac{V_{\ell
L}}{R^{\ell+L+1}}\,, \label
{eq:ap2}
\end{eqnarray}
where
\begin{equation}
V_{\ell L}  = 4\pi (-1)^L (\ell,L)^{-1/2} \sum_{\mu} K_{\ell
L}^{\mu} \;
T^{(\ell)}_{\mu}(\mbox{\boldmath{$\sigma$}})\;
T^{(L)}_{-\mu}(\rhoo) \,. \label {eq:ap3}
\end{equation}
In the above,
$T^{(\ell)}_{\mu}(\mbox{\boldmath{$\sigma$}})$ and
$T^{(L)}_{-\mu}(\rhoo)$  are the atomic
multipole tensor operators defined by
\begin{equation}
T^{(\ell)}_{\mu}(\mbox{\boldmath{$\sigma$}}) = \sum_{i} Q_i
\sigma_i^\ell Y_{\ell \mu}(\hat
{\mbox{\boldmath{$\sigma$}}}_i)\, ,\label{mpol}
\end{equation}
and
\begin{equation}
T^{(L)}_{-\mu}(\rhoo) = \sum_{j} q_j \rho_j^L Y_{L -\mu}(\hat
{\rhoo}_j)\, ,
\end{equation}
and $Q_i$ and $\mbox{\boldmath{$\sigma$}}_i$ are the charge and
the position vector of the $i^{\it th}$ particle in atom $a$,
respectively. Similarly, $q_j$ and $\rhoo_j$ are for
the $j^{\it th}$ particle in atom $b$. The
coefficient $K_{\ell L}^{\mu}$ in Eq. (\ref {eq:ap3}) is
\begin{eqnarray}
K_{\ell L}^{\mu} &=& \left[ { {\ell+L}\choose {\ell+\mu} } {
{\ell+L}\choose {L+\mu}
}\right]^{1/2}\, \label {eq:ap4}
\end{eqnarray}
and $(\ell,L,\cdots)=(2\ell+1)(2L+1)\cdots$.

Since the Coulomb interaction $V$ is cylindrically symmetric about
the molecular axis $\textbf{R}$ or ${z}$
axis~\cite{marinescu97,Singer05}, the projection of the total
angular momentum of the combined system $a$-$b$ along the $z$ axis
(with magnetic quantum number $M$), is conserved. Therefore,
states with $M=\pm2$, $\pm1$, and $0$ are not mixed with each
other, corresponding to the $\Delta$, $\Pi$, and $\Sigma$
molecular states, respectively. The $\Delta$ and $\Pi$ states are
degenerate with respect to the sign of $M$ and the degeneracy can
not be removed physically in the free combined system $a$-$b$.
Therefore, we only study the states with positive $M$ in this
work.
\subsection {$\Delta$ state}
For two like isotope atoms $a$ and $b$ in $P$ symmetry, the
zeroth-order wave function for
the $\Delta$ state of the combined system $a$-$b$ can be written
in the form:
\begin{eqnarray}
\Psi^{(0)}(\Delta,\beta) &=& \frac{\alpha}{\sqrt 2}[\Psi_{n_{\rm
a}}(M_a; \mbox{\boldmath{$\sigma$}}) \Psi_{n_{\rm b}}(M_b;\rhoo)
+\beta \Psi_{n_{\rm a}}(M_a; \rhoo)\Psi_{n_{\rm
b}}(M_b;\mbox{\boldmath{$\sigma$}})]\,,
\label {eq:ap1}
\end{eqnarray}
where $M_a = M_b=1$ are the magnetic quantum numbers, $\alpha$
is the normalization factor, and $\beta$ describes the symmetry
due
to the exchange of two initial states $\Psi_{n_{\rm a}}$ and
$\Psi_{n_{\rm b}}$. If two atoms are both in the same $P$ state,
then $\alpha =\sqrt{2}$ and $\beta = 0$; if they are in
different $P$ states, then $\alpha = 1$ and $\beta = \pm 1$
~\cite{marinescu97}.

According to the perturbation theory, the first-order energy is
\begin{eqnarray} V^{(1)}&=& \langle\Psi^{(0)}(\Delta,\beta)|V|
\Psi^{(0)}(\Delta,\beta)\rangle \,\nonumber \\&=&
-\frac{C_{5}(\Delta,{\beta})}{R^{5}}\,, \label{eq:ap10}
\end{eqnarray}
where, after some angular momentum algebra, one gets (see also
Ref.~\cite{Zyg94}),
\begin{eqnarray}
C_{5}(\Delta,{\beta}) &=& - A_1+\beta A_2\,, \label{eq:ap11} \\
A_1
&=& \frac{4\pi}{25}\langle
\Psi_{n_a}(\mbox{\boldmath{$\sigma$}}) ||\sum_i Q_i \sigma_i^2
Y_{2}(\hat{\mbox{\boldmath{$\sigma$}}}_i)|| \Psi_{n_
a}(\mbox{\boldmath{$\sigma$}})
\rangle\nonumber \\
&&\times \langle \Psi_{n_b}({\mbox{\boldmath{$\rho$}}}) ||\sum_j
q_j \rho_j^2
Y_{2}(\hat{\mbox{\boldmath{$\rho$}}}_j)||
\Psi_{n_b}({\mbox{\boldmath{$\rho$}}}) \rangle\,,
\label {eq:apa} \\
A_2 &=& \frac{4\pi}{25}| \langle
\Psi_{n_a}(\mbox{\boldmath{$\sigma$}}) ||\sum_i Q_i
\sigma_i^2 Y_{2}(\hat{\mbox{\boldmath{$\sigma$}}}_i)|| \Psi_{n_b}(\mbox{\boldmath{$\sigma$}})\rangle|^2\,. \label {eq:apa9}
\end{eqnarray}

The second-order energy is
\begin{eqnarray}
V^{(2)} &=& -{\sum_{n_sn_t}}'\sum_{L_sM_s}\sum_{L_tM_t} \frac{
|\langle\Psi^{(0)}(\Delta)|V|
\chi_{n_s} (L_sM_s;\mbox{\boldmath{$\sigma$}})\omega_{n_t}
(L_tM_t;{\rhoo})\rangle|^2 }
{E_{n_sn_t}-E^{(0)}_{n_an_b}}\,,\ \label{eq:ap8}
\end{eqnarray}
where $ \chi_{n_s} (L_sM_s;\mbox{\boldmath{$\sigma$}})\omega_{n_t}
(L_tM_t;{\rhoo})$
 is an allowed intermediate state with the energy eigenvalue
$E_{n_sn_t}=E_{n_s}+E_{n_t}$, and the prime in the summation
indicates that the terms with $E_{n_sn_t}=E^{(0)}_{n_an_b}$ should
be excluded. Substituting Eqs. (\ref {eq:ap2}) and (\ref {eq:ap1})
into Eq. (\ref {eq:ap8}), we obtain
\begin{eqnarray}
V^{(2)} &=& -{\sum_{n_sn_t}}'\sum_{L_sM_s}\sum_{L_tM_t} \frac{B_1
+\beta B_2 }
{E_{n_sn_t}-E^{(0)}_{n_an_b}}\,, \label{eq:ap9}
\end{eqnarray}
with
\begin{eqnarray}
B_1 &=& |\langle \Psi_{n_{a}}(M_a;\mbox{\boldmath{$\sigma$}})
\Psi_{n_{b}}(
M_b;{\rhoo})|V| \chi_{n_s} (L_sM_s;\mbox{\boldmath{$\sigma$}})
\omega_{n_t} (L_tM_t;{\rhoo})
\rangle|^2 \,, \label {eq:apc10}\\
 B_2 &=& \langle \Psi_{n_{a}}(
M_a;\mbox{\boldmath{$\sigma$}}) \Psi_{n_{b}}(M_b;{\rhoo})|V|
\chi_{n_s}
(L_sM_s;\mbox{\boldmath{$\sigma$}}) \omega_{n_t} (L_tM_t;{\rhoo})
\rangle\nonumber\\
&\times& \langle \Psi_{n_{a}}(M_a;{\rhoo}) \Psi_{n_{b}}(
M_b;\mbox{\boldmath{$\sigma$}})|V| \chi_{n_s}
(L_sM_s;\mbox{\boldmath{$\sigma$}})
\omega_{n_t} (L_tM_t;{\rhoo}) \rangle\,. \label {eq:apc11}
\end{eqnarray}
After applying the Wigner-Eckart theorem, we have
\begin{eqnarray}
{\sum_{n_sn_t}}'\sum_{L_sM_s}\sum_{L_tM_t}\frac{B_1}{E_{n_sn_t}-E^{(0)}_{n_an_b}}
&=&
\sum_{\ell
\ell'LL'}\frac{D_{1}(\ell,L,\ell',L')}{R^{\ell+L+\ell'+L'+2}}\,,
\label {eq:ap12}
\end{eqnarray}
with
\begin{eqnarray}
D_{1}(\ell,L,\ell',L')&=&\frac{1}{2\pi}\sum_{L_sL_t}G_1(\ell,L,\ell',L',L_s,L_t)
F_1(\ell,L,\ell',L',L_s,L_t)\,. \label {eq:apb18}
\end{eqnarray}
In Eq. (\ref {eq:apb18}), $G_1$ is the angular-momentum part and
$F_1$ is the
oscillator strength part. Their expressions are
\begin{eqnarray}
&&G_1(\ell,L,\ell',L',L_s,L_t)=(-1)^{L+L'}(\ell,L,\ell',L')^{1/2}\sum_{\mu}
K_{\ell
L}^{\mu}K_{\ell' L'}^{\mu} {\left(\matrix{L_a&\ell&L_s\cr{-M_a}&\mu&M_a-\mu\cr}\right)}\nonumber\\
&&\times{\left(\matrix{1&L&L_t\cr{-M_b}&-\mu&M_b+\mu\cr}\right)}
{\left(\matrix{1&\ell'&L_s\cr{-M_a}&\mu&M_a-\mu\cr}\right)}
{\left(\matrix{1&L'&L_t\cr{-M_b}&-\mu&M_b+\mu\cr}\right)}\,,
\label {eq:ap19}
\end{eqnarray}
and
\begin{eqnarray}
F_1(\ell,L,\ell',L',L_s,L_t)& =&
\frac{9\pi}{2}\nonumber\\
&\times&\sum_{n_sn_t}
\frac{\bar{g}_{n_s;n_{a}n_{a}}(L_s,1,1,\ell,\ell')\bar{g}_{n_t;n_{b}n_{b}}(L_t,1,1,L,L')
} {(\Delta E_{n_s,n_a}+\Delta E_{n_t,n_b})|\Delta
E_{n_s,n_a}\Delta E_{n_t,b}|}\,, \label {eq:ap20}
\end{eqnarray}
with
\begin{eqnarray}
  \bar{g}_{n_s;n_{a}n_{b}}(L_s,L_1,L_2,\ell,\ell') &=&
\frac{8\pi}{(\ell,\ell')} \frac{\sqrt{|\Delta E_{n_sn_a}\Delta
E_{n_sn_{b}}|}}{\sqrt{(2L_1+1)(2L_2+1)}}\nonumber\\
&\times& \langle\Psi_{n_{a}}(L_1;\mbox{\boldmath{$\sigma$}})
||\sum_i Q_i \sigma_i^{\ell}
Y_{\ell}(\hat{\mbox{\boldmath{$\sigma$}}}_i)||\chi_{n_s}
(L_s;\mbox{\boldmath{$\sigma$}})\rangle\nonumber\\
&\times&\langle\Psi_{n_{b}}(L_2;\mbox{\boldmath{$\sigma$}})
||\sum_i Q_i \sigma_i^{\ell'}
Y_{\ell'}(\hat{\mbox{\boldmath{$\sigma$}}}_i)||\chi_{n_s}
(L_s;\mbox{\boldmath{$\sigma$}})\rangle\,,
 \label{eq:ap16}
\end{eqnarray}
and $\Delta E_{n_sn_a}=E_{n_s}-E_{n_{a}}$, {\it etc}. For the
special case where the two initial states $\Psi_{n_a}$ and
$\Psi_{n_b}$ are the same and $\ell=\ell'$,
$\bar{g}_{n_s;n_{a}n_{a}}$ reduces to the absolute value of the
$2^\ell$-pole oscillator strength
\begin{eqnarray}
\bar{f}_{n_sn_{a}}^{\ell} &=& \frac{8\pi\Delta
E_{n_sn_a}}{(2\ell+1)^2(2L_1+1)}|\langle\Psi_{n_{a}}(L_1;\mbox{\boldmath{$\sigma$}})
||\sum_i Q_i \sigma_i^{\ell}
Y_{\ell}(\hat{\mbox{\boldmath{$\sigma$}}}_i)||\chi_{n_s}
(L_s;{\mbox{\boldmath{$\sigma$}}})\rangle|^2\,.
 \label{eq:ap17}
\end{eqnarray}
 Similarly, we have
 \begin{eqnarray}
&&{\sum_{n_sn_t}}'\sum_{L_sM_s}\sum_{L_tM_t}\frac{B_2}{E_{n_sn_t}-E^{(0)}_{n_an_b}}
=
\sum_{\ell
\ell'LL'}\frac{D_2(\ell,L,\ell',L')}{R^{\ell+L+\ell'+L'+2}}\,,
\label {eq:apa12}\\
&&
D_{2}(\ell,L,\ell',L')=\frac{1}{2\pi}\sum_{L_sL_t}G_2(\ell,L,\ell',L',L_s,L_t)
F_2(\ell,L,\ell',L',L_s,L_t)\,, \label {eq:ap18}
\end{eqnarray}
with
\begin{eqnarray}
&&G_2(\ell,L,\ell',L',L_s,L_t)
=(-1)^{L+L'}(\ell,L,\ell',L')^{1/2}\sum_{\mu}
K_{\ell L}^{\mu}K_{\ell' L'}^{M_b-M_a+\mu}\nonumber\\
&&\times
{\left(\matrix{1&\ell&L_s\cr{-M_a}&\mu&M_a-\mu\cr}\right)}
{\left(\matrix{1&L'&L_t\cr{-M_a}&M_a-M_b-\mu&M_b+\mu\cr}\right)}\nonumber\\
&&\times {\left(\matrix{1&L&L_t\cr{-M_b}&-\mu&M_b+\mu\cr}\right)}
{\left(\matrix{1&\ell'&L_s\cr{-M_b}&M_b-M_a+\mu&M_a-\mu\cr}\right)}\,,
\label {eq:ap23}
\end{eqnarray}
\begin{eqnarray}
&& F_2(\ell,L,\ell',L',L_a,L_b,L_s,L_t) = \nonumber\\
&&  \frac{9 \pi}{2} \sum_{n_sn_t}
\frac{\bar{g}_{n_s;n_{a}n_b}(L_s,1,1,\ell,\ell')\bar{g}_{n_t;n_{a}n_b}(L_t,1,1,L',L)
} {(\Delta E_{n_s,n_a}+\Delta E_{n_t,n_b})\sqrt{|\Delta
E_{n_s,n_a}\Delta E_{n_s,n_b}\Delta E_{n_t,n_a}\Delta
E_{n_t,n_b}}|}\,. \label {eq:ap24}
\end{eqnarray}
Finally, the second-order
energy $V^{(2)}$  is
\begin{eqnarray}
V^{(2)} &=&  -\sum_{n\geq 3}
\frac{C_{2n}(\Delta,\beta)}{R^{2n}}\,, \label {eq:ap25}
\end{eqnarray}
where the dispersion coefficients $C_{2n}(\Delta,\beta)$ are
defined by
\begin{eqnarray}
C_{2n}(\Delta,\beta) &=& \mathop{\sum_{\ell,L,\ell',L'\ge
1}}_{\ell+L+\ell'+L'+2=2n} [D_{1}(\ell,L,\ell',L')+\beta
D_{2}(\ell,L,\ell',L')]\,. \label {eq:ap26}
\end{eqnarray}
\subsection {$\Pi$ state}
For the $\Pi$ state, the zeroth-order wave function is in the
following form
\begin{eqnarray}
\Psi^{(0)}(\Pi,\beta,\gamma)&=& \frac{\alpha}{2}[\Psi_{n_a}(
M_a;\mbox{\boldmath{$\sigma$}}) \Psi_{n_
b}(M_b;{\bf\rhoo})+\gamma\Psi_{n_
a}(M_b;\mbox{\boldmath{$\sigma$}}) \Psi_{n_b}(M_a;{\bf \rhoo})] \nonumber \\
&+&\frac{\alpha\beta}{2}[\Psi_{n_a}(M_a;{\bf \rhoo})\Psi_{n_
b}(M_b;\mbox{\boldmath{$\sigma$}})+\gamma\Psi_{n_a}(M_b;{\bf
\rhoo})\Psi_{n_
b}(M_a;\mbox{\boldmath{$\sigma$}})]\,, \label {eq:ap34}
\end{eqnarray}
where $\beta=\pm 1$, $M_a$=0, and $M_b=1$. If $n_a=n_b$,
$\alpha = \sqrt{2}$ and $\gamma = 0$. If $n_a\neq n_b$,
then $\alpha = 1$ and $\gamma = \pm 1$. The first-order
energy correction yields
\begin{eqnarray}
C_{5}(\Pi,{\beta},\gamma)
&=&-\alpha^2(1+2\gamma+\gamma^2)C_{5}(\Delta,{\beta}).
\label{eq:ap32}
\end{eqnarray}
The second-order energy correction is
\begin{eqnarray}
V^{(2)}(\Pi,\beta,\gamma) &=&
-{\sum_{n_sn_t}}'\sum_{L_sM_s}\sum_{L_tM_t} \frac{
|\langle\Psi^{(0)}(\Pi,\beta,\gamma)|V| \chi_{n_s} (L_sM_s;{\bf
r})\omega_{n_t} (L_tM_t;{\bf
\rho})\rangle|^2 } {E_{n_sn_t}-E^{(0)}_{n_an_b}}\,.
\label{eq:ap48}
\end{eqnarray}
Introducing a function $g_o$ defined by
\begin{eqnarray}
&& g_o(M_a,M_b,M_c,M_d,\ell,L,L_s,L_t,\ell',L') =(-1)^{L+L'}
(\ell,L,\ell',L')^{1/2}\sum_{\mu\mu'M_sM_t} K_{\ell L}^{\mu}K_{\ell'L'}^{\mu'}\nonumber\\
&& \times {\left(\matrix{1&\ell&L_s\cr{-M_a}&\mu&M_s\cr}\right)}
{\left(\matrix{1&L&L_t\cr{-M_b}&-\mu&M_t\cr}\right)}
 {\left(\matrix{1&\ell'&L_s\cr{-M_c}&\mu'&M_s\cr}\right)}
{\left(\matrix{1&L'&L_t\cr{-M_d}&-\mu'&M_t\cr}\right)}\,, \label
{eq:ap49}
\end{eqnarray}
one can write $V^{(2)}(\Pi,\beta,\gamma)$ as
\begin{eqnarray}
V^{(2)}(\Pi,\beta,\gamma) =-\sum_{n\geq 3}
\frac{C_{2n}(\Pi,\beta,\gamma)}{R^{2n}}\,,
\end{eqnarray}
where
\begin{eqnarray}
C_{2n}(\Pi,\beta,\gamma)= \mathop{\sum_{\ell,L,\ell',L'\ge
1}}_{\ell+L+\ell'+L'+2=2n}
D_{\beta\gamma}(\ell,L,\ell',L') \,,
\end{eqnarray}
\begin{eqnarray}
&&D_{\beta\gamma}(\ell,L,\ell',L')=
\frac{9\alpha^2}{16}\sum_{L_sL_t}\bigg[f_{1\gamma}(\ell,L,L_s,L_t,\ell',L')\;
G_1 + f_{2\gamma}(\ell,L,L_s,L_t,\ell',L')\;G_2 \bigg]\,,\label
{eq:kap10}
\end{eqnarray}
\begin{eqnarray}
&&f_{1\gamma}(\ell,L,L_s,L_t,\ell',L')=  \nonumber\\
&&\frac{1}{2}(1+\gamma^2)[g_o(0,1,0,1,\ell,L,L_s,L_t,\ell',L')+g_o(1,0,1,0,\ell,L,L_s,L_t,\ell',L')]\nonumber\\
&&+\gamma\,g_o(0,1,1,0,\ell,L,L_s,L_t,\ell',L')+\gamma\,g_o(1,0,0,1,\ell,L,L_s,L_t,\ell',L')\,,\label
{eq:kap5}
\end{eqnarray}
\begin{eqnarray}
&&f_{2\gamma}(\ell,L,L_s,L_t,\ell',L')= \nonumber\\
&&\frac{1}{2}(1+\gamma^2)[g_o(0,1,1,0,\ell,L,L_s,L_t,\ell',L')+g_o(1,0,0,1,\ell,L,L_s,L_t,\ell',L')]\nonumber\\
&&+\gamma\,g_o(0,1,0,1,\ell,L,L_s,L_t,\ell',L')+\gamma\,g_o(1,0,1,0,\ell,L,L_s,L_t,\ell',L')\,,\label
{eq:kap6}
\end{eqnarray}
\begin{eqnarray}
G_1={\sum_{n_sn_t}}'\bigg[
\frac{\bar{g}_{n_s;n_{a}n_a}(L_s,1,1,\ell,\ell')\bar{g}_{n_t;n_{b}n_b}(L_t,1,1,L,L')
}
{(\Delta E_{n_s,n_a}+\Delta E_{n_t,n_b})|\Delta E_{n_s,n_a}\Delta
E_{n_t,n_b}|}\nonumber\\
+
\frac{\bar{g}_{n_s;n_{b}n_b}(L_s,1,1,\ell,\ell')\bar{g}_{n_t;n_{a}n_a}(L_t,1,1,L,L')
}
{(\Delta E_{n_s,n_a}+\Delta E_{n_t,n_b})|\Delta E_{n_s,n_a}\Delta
E_{n_t,n_b}|}\bigg]\,,
\label{eq:wap55}
\end{eqnarray}
and
\begin{eqnarray}
G_2=\beta {\sum_{n_sn_t}}'\,\bigg[
\frac{\bar{g}_{n_s;n_{a}n_b}(L_s,1,1,\ell,\ell')\bar{g}_{n_t;n_{b}n_a}(L_t,1,1,L,L')
}
{(\Delta E_{n_s,n_a}+\Delta E_{n_t,n_b})\sqrt{|\Delta
E_{n_s,n_a}\Delta E_{n_s,n_b}\Delta E_{n_t,n_a}\Delta E_{n_t,n_b}|}}\nonumber\\
 +\frac{\bar{g}_{n_s;n_{b}n_a}(L_s,1,1,\ell,\ell')\bar{g}_{n_t;n_{a}n_b}(L_t,1,1,L,L') }
{(\Delta E_{n_s,n_a}+\Delta E_{n_t,n_b})\sqrt{|\Delta
E_{n_s,n_a}\Delta E_{n_s,n_b}\Delta
E_{n_t,n_a}\Delta E_{n_t,n_b}|}}\bigg]\,. \label{eq:wap56}
\end{eqnarray}
From the above expressions, it is clear that $f_{1\gamma}=\gamma
f_{2\gamma}$ when
$\gamma=\pm 1$ and $G_2 = \beta G_1$ when $\gamma=0$.
\subsection {$\Sigma$ state}
For the $\Sigma$ state, the possible zeroth-order wave functions
for the combined system $a$-$b$
are
\begin{eqnarray}
\Psi^{(0)}(\Sigma,\beta) &=& \frac{\alpha}{\sqrt
2}[\Psi_{n_a}(M_a;\mbox{\boldmath{$\sigma$}})
\Psi_{n_b}(M_b;{\bf \rho}) +\beta \Psi_{n_a}(M_a;{\bf
\rho})\Psi_{n_b}(M_b;\mbox{\boldmath{$\sigma$}})]\,, \label
{eq:ap33}
\end{eqnarray}
and
\begin{eqnarray}
\Psi^{(0)}(\Sigma,\beta,\gamma) =
\frac{\alpha}{2}[\Psi_{n_a}(M_a;\mbox{\boldmath{$\sigma$}})
\Psi_{n_b}(M_b;{\bf\rho})+\gamma\Psi_{n_a}(M_b;\mbox{\boldmath{$\sigma$}}) \Psi_{n_b}(M_a;{\bf \rho})] \nonumber \\
+\frac{\alpha\beta}{ 2}[\Psi_{n_a}(M_a;{\bf
\rho})\Psi_{n_b}(M_b;\mbox{\boldmath{$\sigma$}})+\gamma\Psi_{n_a}(M_b;{\bf
\rho})\Psi_{n_b}(M_a;\mbox{\boldmath{$\sigma$}})]\,. \label
{eq:apa34}
\end{eqnarray}
In the $\Psi^{(0)}(\Sigma,\beta)$ state we have $M_a=M_b=0$, while
in the $\Psi^{(0)}(\Sigma,\beta,\gamma)$ state, $M_a = -M_b=1$,
and
$\gamma=\pm 1$. When $n_a = n_b$, $\beta = 0$ and $\alpha = \sqrt
2$,
while otherwise $\beta = \pm 1$ and $\alpha = 1$.
For the $\Sigma$ state, there is an
additional symmetry with respect to the reflection of wave
function on a plane containing the molecular axis. If $Q$ is the
reflection operator~\cite{marinescu97},
\begin{eqnarray}
Q \Psi_{n_a}(M_a;{\bf r})=(-1)^{M_a}\Psi_{n_a}(-M_a;{\bf r})
\,, \label {eq:ap35}
\end{eqnarray}
then
\begin{eqnarray}
Q \Psi^{(0)}(\Sigma,\beta) =\Psi^{(0)}(\Sigma,\beta) \,, \label
{eq:ap36}
\end{eqnarray}
and
\begin{eqnarray}
Q \Psi^{(0)}(\Sigma,\beta,\gamma)
=\gamma\,\Psi^{(0)}(\Sigma,\beta,\gamma) \,. \label
{eq:ap37}
\end{eqnarray}
Due to this symmetry, the state $\Psi^{(0)}(\Sigma,\beta,-1)$ can
not be mixed with other
two states by the Coulomb interaction $V$. However, the degenerate
perturbation theory is
required to remove the degeneracy between
$\Psi^{(0)}(\Sigma,\beta)$ and
$\Psi^{(0)}(\Sigma,\beta,1)$.
\subsubsection {The first-order energy}
For the $\Psi^{(0)}(\Sigma,\beta,-1)$ state, the first-order
energy is
\begin{eqnarray}
\langle \Psi^{(0)}(\Sigma,\beta,-1)|V|
\Psi^{(0)}(\Sigma,\beta,-1)\rangle
&=&0. \label {eq:ap38}\nonumber\\
\end{eqnarray}
 For $\Psi^{(0)}(\Sigma,\beta)$ and
$\Psi^{(0)}(\Sigma,\beta,1)$, the matrix elements of the Coulomb
interaction $V$ are
\begin{eqnarray}
\langle \Psi^{(0)}(\Sigma,\beta)|V|
\Psi^{(0)}(\Sigma,\beta)\rangle=-\frac{4
C_5(\Delta,\beta)}{R^5}\,,  \end{eqnarray}
\begin{eqnarray}
\langle \Psi^{(0)}(\Sigma,\beta)|V|
\Psi^{(0)}(\Sigma,\beta,1)\rangle = -\frac{2\sqrt2
C_5(\Delta,\beta)}{R^5}, \label {eq:ap40}
\end{eqnarray}
and
\begin{eqnarray} \langle \Psi^{(0)}(\Sigma,\beta,1)|V|
\Psi^{(0)}(\Sigma, \beta, 1)\rangle =-\frac{2
C_5(\Delta,\beta)}{R^5}\,.
\end{eqnarray}
The first-order energies are obtained by the diagonalization of
the following matrix:
\begin{eqnarray}
 -\frac{4
C_5(\Delta,\beta)}{R^5} {\left[\matrix{1&\frac{1}{\sqrt
2}\cr{\frac{1}{\sqrt
2}}&\frac{1}{2}\cr}\right]}\,.
\end{eqnarray}
They are
\begin{eqnarray}
&&\lambda_1=0\,, \\
&&\lambda_2=-\frac{6 C_5(\Delta,\beta)}{R^5}\,.
\end{eqnarray}
Their corresponding normalized eigenvectors are
\begin{eqnarray}
&&\Psi^{(0)}(\Sigma,\beta; \lambda_1)=-\frac{1}{\sqrt
3}\Psi^{(0)}(\Sigma,\beta)+\sqrt\frac{2}{
3}\Psi^{(0)}(\Sigma,\beta,1)\,, \\
&&\Psi^{(0)}(\Sigma,\beta; \lambda_2)=\sqrt\frac{2}{
3}\Psi^{(0)}(\Sigma,\beta)+\sqrt\frac{1}{
3}\Psi^{(0)}(\Sigma,\beta,1)\,,
\end{eqnarray}
where we introduce the semicolon to denote the use of the label
$\lambda$ now. Thus, $C_5$ for $\Psi^{(0)}(\Sigma,\beta;
\lambda_1)$ is zero and $C_5$ for $\Psi^{(0)}(\Sigma,\beta;
\lambda_2)$ is
\begin{eqnarray}
 C_5(\Sigma,\beta; \lambda_2)=6\,C_5(\Delta,\beta).
\end{eqnarray}
\subsubsection{The second-order energy}
To be convenient, we use the expression $\Psi^{(0)}(\Sigma,\beta;
\lambda_3)$ instead of $\Psi^{(0)}(\Sigma,\beta,-1)$. For the
state $\Psi^{(0)}(\Sigma,\beta; \lambda_i)$, the second-order
energy correction can be written in the form
\begin{eqnarray}
V^{(2)}(\Sigma,\beta; \lambda_i) =-\sum_{n\geq 3}
\frac{C_{2n}(\Sigma,\beta; \lambda_i)}{R^{2n}}\,, \label {eq:ap51}
\end{eqnarray}
where
\begin{eqnarray}
&& C_{2n}(\Sigma,\beta; \lambda_i)=
\mathop{\sum_{\ell,L,\ell',L'\ge
1}}_{\ell+L+\ell'+L'+2=2n} D_{\lambda_i}(\ell,L,\ell',L') \,, \label {eq:ap52}\\
&&D_{\lambda_i}(\ell,L,\ell',L')=\frac{3\alpha^2}{8}\sum_{L_sL_t}f_{\lambda_i}(\ell,L,L_s,L_t,\ell',L')
(G_1+(2-i)^{(2-i)}G_2)\,. \label {eq:ap53}
\end{eqnarray}
and $f_{\lambda_i}$ are
\begin{eqnarray}
 &&f_{\lambda_1}(\ell,L,L_s,L_t,\ell',L')=g_o(0,0,0,0,\ell,L,L_s,L_t,\ell',L') \,\\
&& -g_o(0,0,-1,1,\ell,L,L_s,L_t,\ell',L')- g_o(0,0,1,-1,\ell,L,L_s,L_t,\ell',L')\nonumber\\
&&-g_o(-1,1,0,0,\ell,L,L_s,L_t,\ell',L')+ g_o(-1,1,-1,1,\ell,L,L_s,L_t,\ell',L')\nonumber\\
&&+g_o(-1,1,1,-1,\ell,L,L_s,L_t,\ell',L')- g_o(1,-1,0,0,\ell,L,L_s,L_t,\ell',L')\nonumber\\
&&+g_o(1,-1,-1,1,\ell,L,L_s,L_t,\ell',L')+
g_o(1,-1,1,-1,\ell,L,L_s,L_t,\ell',L')\,,\nonumber
\end{eqnarray}
\begin{eqnarray}
&&f_{\lambda_2}(\ell,L,L_s,L_t,\ell',L')=g_o(0,0,0,0,\ell,L,L_s,L_t,\ell',L')\, \label {eq:ap60}\\
&&+2g_o(0,0,-1,1,\ell,L,L_s,L_t,\ell',L')+g_o(0,0,1,-1,\ell,L,L_s,L_t,\ell',L')\nonumber\\
&&+g_o(-1,1,0,0,\ell,L,L_s,L_t,\ell',L')+\frac{1}{2}\, g_o(-1,1,-1,1,\ell,L,L_s,L_t,\ell',L')\nonumber\\
&&+\frac{1}{2}\, g_o(-1,1,1,-1,\ell,L,L_s,L_t,\ell',L')+g_o(1,-1,0,0,\ell,L,L_s,L_t,\ell',L')\nonumber\\
&&
+\frac{1}{2}g_o(1,-1,-1,1,\ell,L,L_s,L_t,\ell',L')+\frac{1}{2}\,
g_o(1,-1,1,-1,\ell,L,L_s,L_t,\ell',L')\,, \nonumber
\end{eqnarray}
and
\begin{eqnarray}
&& f_{\lambda_3}(\ell,L,L_s,L_t,\ell',L')=\label {eq:kap7}\\
&&\frac{3}{2}
[\,g_o(-1,1,-1,1,\ell,L,L_s,L_t,\ell',L')-g_o(-1,1,1,-1,\ell,L,L_s,L_t,\ell',L')\nonumber\\
&&-g_o(1,-1,-1,1,\ell,L,L_s,L_t,\ell',L')+g_o(1,-1,1,-1,\ell,L,L_s,L_t,\ell',L')]\,.\nonumber
\end{eqnarray}
\section {Calculations and Results}
 In the center-of-mass frame, the Hamiltonian of a helium atom can be written in the form
\begin{eqnarray}
H &=& - \frac{1}{2\mu_e}\nabla_{{\bf r}_1}^2 -
\frac{1}{2\mu_e}\nabla_{{\bf r}_2}^2
-\frac{1}{m_n}\nabla_{{\bf r}_1}\cdot\nabla_{{\bf r}_2}
-\frac{2}{r_1}-\frac{2}{r_2}
+\frac{1}{r_{12}}\,, \label {eq:t20}
\end{eqnarray}
where $m_n$ is the nucleus mass, $\mu_e$ is the reduced mass
between the electron and the
nucleus, $\textbf{r}_1$ and $\textbf{r}_2$ are the position
vectors of the two electrons
relative to the nucleus, and $r_{12}$ is the distance between
them. To calculate dispersion
coefficients $C_n$, we variationally evaluate energy eigenvalues
and corresponding
eigenfunctions of the initial states He($2\,^1\!P$) and
He($2\,^3\!P$) with the correlated
Hylleraas basis set
\begin{eqnarray}
 \{r_1^i\,r_2^j\,r_{12}^k\,e^{-\alpha r_1-\beta r_2}
{\cal Y}_{{\ell_1}{\ell_2}}^{{LM}}({\hat{\bf r}}_1,{\hat{\bf
r}}_2)\} \,,
\end{eqnarray}
where ${\cal Y}_{{\ell_1}{\ell_2}}^{{LM}}({\hat{\bf
r}}_1,{\hat{\bf
r}}_2)$ is the coupled spherical harmonics for the two electrons
forming a common eigenstate of ${\bf L}^2$ and $L_z$, and $\alpha$
and
$\beta$ are two nonlinear parameters, which are optimized by first
calculating analytically the first-order derivatives of the
variation
energy with respect to these parameters and then using Newton's
method to find their roots.
The basis set includes all combinations of $i\geq \ell_1$, $j\geq
\ell_2$, and $k \geq 0$ with $i+j+k\leq \Omega$, where $\Omega$ is
an integer controlling the size of the basis set. Then we generate
the spectra of the intermediate states for the $S$, $P$, $D$, $F$,
and $G$ symmetries by diagonalizing directly the Hamiltonian of
the helium atom in chosen basis sets. In addition, we need to
transform the $2^{\ell}$-pole transition operator
$T^{\ell}_{\mu}(\mbox{\boldmath{$\sigma$}})$, defined in Eq.
(\ref{mpol}), into the center-of-mass coordinates and calculate
the reduced matrix elements of dipole, quadrupole, octupole, and
hexapole transition operators (for details, refer to Ref.~\cite
{zhy04}).

Table~\ref{a1} shows the convergence study of the nonrelativistic
energy of the $^\infty\!\,$He($2\,^3\!P$) state with the increase
of the size of basis set. Compared to the value of Drake {\it et
al.}~\cite{drake02}, our result is accurate to about 15 digits.
Table~\ref{b1} gives the convergence pattern of
$C_{6}({\Delta,0})$, $C_{6}({\Pi,+,0})$, and $C_{6}({\Sigma,0;
\lambda_1})$ for
$^\infty\!\,$He($2\,^3\!P$)--$^\infty\!\,$He($2\,^3\!P$) as the
sizes of basis sets, including the initial state and the three
intermediate states, increase progressively.

Table~\ref{g1} presents the long-range interaction coefficients
$C_{5}$, $C_{6}$, $C_{8}$, and $C_{10}$ for the
He($2\,^1\!P$)-He($2\,^1\!P$) system. Table~\ref{g2} shows
$C_{5}$, $C_{6}$, $C_{8}$, and $C_{10}$ for the
He($2\,^1\!P$)-He($2\,^3\!P$) system. Table~\ref{g3} lists
$C_{5}$, $C_{6}$, $C_{8}$, and $C_{10}$ for the
He($2\,^3\!P$)-He($2\,^3\!P$) system. We note that, for the three
He($n\,P$)--He$(n'\,P)$ systems, $C_{5}(\Pi,-,\beta)$,
$C_5(\Sigma,\beta; \lambda_1)$, and $C_5(\Sigma,\beta; \lambda_3)$
 are zero, $C_{5}(\Pi,+,\beta)$ and $C_6$ are positive, and both $C_{5}(\Delta,\beta)$
 and $C_5(\Sigma,\beta; \lambda_2)$ are negative.
 For the $\Psi^{(0)}(\Sigma,\beta; \lambda_3)$ states, $C_8$ and $C_{10}$ are
negative in the three tables.

Ovsyannikov obtained expressions for the $C_6$ (dispersion)
coefficients between two excited atoms~\cite{Ovs82} and evaluated
the  He($2\,^3\!P_2$)--He($2\,^3\!P_2$)  coefficients using the
atomic dynamic polarizability obtained with a model potential for
the  He($2\,^3\!P$)  atom. The diagonal elements of the long-range
interaction were given in the $jj$ representation; after
transformation to the $LS$ representation~\cite{Yur02} the results
of Ref.~\cite{Ovs82} can be compared with ours. The one
significant discrepancy is between our results for the summed
$C_6(\Sigma, 0;\lambda)$ and that of Ref. \cite{Ovs82}. In
table~\ref{compare} we compare the transformed coefficients of
Ref.~\cite{Ovs82} with the present calculations.

To the best of our knowledge, there are no other published results
for the dispersion coefficients for interaction between
He($2\,^1\!P$)-He($2\,^1\!P$) and He($2\,^1\!P$)-He($2\,^3\!P$).
\acknowledgments This work is supported by the Natural Sciences
and Engineering Research Council of Canada, by the Canadian
computing facilities of ACEnet, SHARCnet, and WestGrid, by DOE and
by NSF through a grant for the Institute of Theoretical Atomic,
Molecular and Optical Physics (ITAMP) at Harvard University and
Smithsonian Astrophysical Observatory. JYZ and ZCY would like to
thank ITAMP for its hospitality during their visits. ZCY would
also like to acknowledge the support by NSC of ROC during his
visit at the Institute of Atomic and Molecular Sciences, Academia
Sinica.

\newpage
\begin{longtable}{ccl}
\caption{\label{a1} Convergence study for the nonrelativistic
energy of He($2\,^3\!P$) with the infinite nuclear mass. $N$
denotes the number of terms in the
basis set. Units are atomic units.}\\
\hline\hline \multicolumn{1}{c}{$\Omega$}&
\multicolumn{1}{c}{$N$}& \multicolumn{1}{c}{$E(\Omega)$}\\
\hline
 12 & 910& --2.133\,164\,190\,779\,194 \\
 13 & 1120&--2.133\,164\,190\,779\,246 \\
 14 & 1360& --2.133\,164\,190\,779\,279\\
 15 & 1632& --2.133\,164\,190\,779\,281\,2\\
 16 & 1892& --2.133\,164\,190\,779\,282\,6\\
Drake~\cite{drake02}& &--2.133\,164\,190\,779\,283\,202(5)\\
 \hline\hline
\end{longtable}
\begin{longtable}{c c c c c c c}
\caption{\label{b1} Convergence study of $C_{6}({\Delta,0})$,
$C_{6}({\Pi,+,0})$, and $C_{6}({\Sigma,0; \lambda_1})$, in atomic
units, for
$^\infty\!\,$He($2\,^3\!P$)--$^\infty\!\,$He($2\,^3\!P$).
 $N_{2\,^3\!P}$, $N_{\,^3\!S}$, $N_{(pp)\,^3\!P}$,
and $N_{\,^3\!D}$ denote respectively the sizes of bases for the
initial state and the three intermediate states of symmetries
$\,^3\!S$, $(pp)\,^3\!P$, and $\,^3\!D$. }\\ \hline\hline
\multicolumn{1}{c}{$N_{2\,^3\!P}$}&
\multicolumn{1}{c}{$N_{\,^3\!S}$}&
\multicolumn{1}{c}{$N_{(pp)\,^3\!P}$}&
\multicolumn{1}{c}{$N_{\,^3\!D}$}&
\multicolumn{1}{c}{$C_{6}({\Delta,0})$}&
\multicolumn{1}{c}{$C_{6}({\Pi,+,0})$}&
\multicolumn{1}{c}{$C_{6}({\Sigma,0; \lambda_1})$}\\
\hline
   1360 &560 & 1230 &853 &4083.026\,993\,80 &2324.937\,567\,24&7296.821\,324\,76    \\
1632 &680 & 1430 &1071 &4083.026\,997\,99     &2324.937\,572\,24&7296.821\,330\,48\\
1938 &816 & 1650 &1323 &4083.026\,998\,66      &2324.937\,572\,91&7296.821\,331\,49\\
\hline\hline
\end{longtable}
\newpage
\begin{longtable}{l c c c}
\caption{\label{g1} The long-range interaction coefficients,
 in
atomic units, for the
He($2\,^1\!P$)-He($2\,^1\!P$) system .
}\\
\hline\hline \multicolumn{1}{l}{$C_n$}&
\multicolumn{1}{l}{$^\infty\!\,$He($2\,^1\!P$)--$^\infty\!\,$He($2\,^1\!P$)}&
\multicolumn{1}{l}{$^4\!\,$He($2\,^1\!P$)--$^4\!\,$He($2\,^1\!P$)}& \multicolumn{1}{l}{$^3\!\,$He($2\,^1\!P$)--$^3\!\,$He($2\,^1\!P$)}\\
\hline
 $C_{5}(\Delta,0)$&$-227.26160604(2)$&$-227.31888382(1)$&$-227.33762037(2)$\\
$ C_{5}(\Pi,+,0)$&909.04642409(1)&909.27553527(2)&909.35048146(3)\\
$C_{5}(\Pi,-,0)$&0&0&0\\
$C_5(\Sigma,0; \lambda_1)$&0&0&0\\
$C_5(\Sigma,0; \lambda_2)$&$-1363.56963613(1)$&$-1363.91330289(1)$&$-1364.02572224(5)$\\
$C_5(\Sigma,0; \lambda_3)$&0&0&0\\
 $C_{6}(\Delta,0)$&5578.63(2)&5584.99(1)&5587.07(1)\\
$ C_{6}(\Pi,+,0)$&3606.63(2)&3610.56(2)&3611.84(1) \\
$ C_{6}(\Pi,-,0)$&10831.74(2) &10844.25(2)&10848.35(2) \\
$C_{6}(\Sigma,0; \lambda_1)$&9682.78(1)&9693.98(2)&9697.63(2)\\
$C_{6}(\Sigma,0; \lambda_2)$&1025.14(2)&1032.23(2)&1034.58(3) \\
$C_{6}(\Sigma,0; \lambda_3)$&1966.062(1)&1968.141(1)& 1968.821(1)\\
$C_{8}(\Delta,0)$&1556.9(6)&1458.9(7)&1426.6(6)\\
$ C_{8}(\Pi,+,0)$&$-170489.4(5) $&$-170548.7(5)$&$-170567.9(3)$\\
$ C_{8}(\Pi,-,0)$&1315208(3)&1316393(2)& 1316781(2) \\
$C_{8}(\Sigma,0; \lambda_1)$&1106570(2)&1107553(2)&1107874(2) \\
$C_{8}(\Sigma,0; \lambda_2)$&1788288(2)&1790861(2)&1791704(2) \\
$C_{8}(\Sigma,0; \lambda_3)$ &$-16971.98(4)$&$-17030.68(3)$&$-17049.91(3)$\\
$C_{10}(\Delta,0)$&$3.61343(3)\times10^{6}$&$3.61541(2)\times10^{6}$&$3.61607(3)\times10^{6}$ \\
$ C_{10}(\Pi,+,0)$&$5.163864(4)\times10^{7}$&$5.168052(3)\times10^{7}$&$5.169423(3)\times10^{7}$\\
$ C_{10}(\Pi,-,0)$&$1.7334869(4)\times10^{8}$&$1.7349273(2)\times10^{8}$& $1.7353990(4)\times10^{8}$\\
$C_{10}(\Sigma,0; \lambda_1)$&$2.0453997(5)\times10^{8}$&$2.0471586(3)\times10^{8}$&$2.0477345(5)\times10^{8}$\\
$C_{10}(\Sigma,0; \lambda_2)$&$7.6324701(1)\times10^{8}$&$7.6394035(3)\times10^{8}$&$7.6416728(2)\times10^{8}$\\
$C_{10}(\Sigma,0; \lambda_3)$&$-2.992224(2)\times10^{6}$&$-2.996113(3)\times10^{6}$&$-2.997383(2)\times10^{6}$ \\
 \hline\hline
\end{longtable}
\newpage
\begin{longtable}{l c c c}
\caption{\label{g2} The long-range interaction coefficients,
 in atomic units, for the
He($2\,^1\!P$)-He($2\,^3\!P$) system .
}\\
\hline\hline \multicolumn{1}{l}{Mass}&
\multicolumn{1}{l}{$^\infty\!\,$He($2\,^1\!P$)--$^\infty\!\,$He($2\,^3\!P$)}&
\multicolumn{1}{l}{$^4\!\,$He($2\,^1\!P$)--$^4\!\,$He($2\,^3\!P$)}& \multicolumn{1}{l}{$^3\!\,$He($2\,^1\!P$)--$^3\!\,$He($2\,^3\!P$)}\\
\hline
 $C_{5}(\Delta,\pm)$&$-189.519232605(3)$&$-189.513470009(3)$&$-189.511579458(2)$\\
$ C_{5}(\Pi,+,\pm)$&758.07693042(3)&758.05388004(2)&758.04631785(3)\\
$C_{5}(\Pi,-,\pm)$&0&0&0\\
$C_5(\Sigma,\pm; \lambda_1)$&0&0&0\\
$C_5(\Sigma,\pm; \lambda_2)$&$-1137.11539564(3)$&$-1137.08082006(2)$&$-1137.06947676(2)$\\
$C_5(\Sigma,\pm; \lambda_3)$&0&0&0\\
 $C_{6}(\Delta,\pm)$&5186.93(2)&5190.06(3)&5191.91(2)\\
$ C_{6}(\Pi,+,\pm)$&2990.19(1)&2992.03(2)&2992.640(2) \\
$ C_{6}(\Pi,-,\pm)$& 10353.76(2)&10359.97(5)&10362.05(2) \\
$C_{6}(\Sigma,\pm; \lambda_1)$&9245.11(2)&9250.65(3)&9252.52(3)\\
$C_{6}(\Sigma,\pm; \lambda_2)$&6162.84(4)&6165.27(3)&6166.13(4) \\
$C_{6}(\Sigma,\pm; \lambda_3)$&1505.135(1)&1506.081(1)&1506.390(1) \\
$C_{8}(\Delta,\pm)$&$-41757.6(6)$&$-41832.7(4)$&$-41856.3(4)$\\
$ C_{8}(\Pi,+,\pm)$&$ -88728.8(2)$&$-88801.9(1)$&$-88825.1(2)$\\
$ C_{8}(\Pi,-,\pm)$&993884(3)&994253(2)& 994375(2) \\
$C_{8}(\Sigma,\pm; \lambda_1)$&843530(3)&843832(2)&843931(2) \\
$C_{8}(\Sigma,\pm; \lambda_2)$&2139396(4)&2140392(3)&2140712(2) \\
$C_{8}(\Sigma,\pm; \lambda_3)$ &$-11571.12(3)$&$-11605.17(2)$&$-11616.32(3)$\\
$C_{10}(\Delta,\pm)$&$2.32333(4)\times10^{6}$&$2.32313(3)\times10^{6}$&$2.32312(1)\times10^{6}$ \\
$ C_{10}(\Pi,+,\pm)$&$5.236817(3)\times10^{7}$&$5.238065(4)\times10^{7}$&$5.238463(1)\times10^{7}$\\
$ C_{10}(\Pi,-,\pm)$&$1.224688(2)\times10^{8}$&$1.224994(2)\times10^{8}$&$1.225094(1)\times10^{8}$ \\
$C_{10}(\Sigma,\pm; \lambda_1)$&$1.435423(2)\times10^{8}$&$1.435806(2)\times10^{8}$&$1.435933(1)\times10^{8}$\\
$C_{10}(\Sigma,\pm; \lambda_2)$&$5.737754(2)\times10^{8}$&$5.739452(2)\times10^{8}$&$5.740007(2)\times10^{8}$\\
$C_{10}(\Sigma,\pm; \lambda_3)$&$-2.018823(3)\times10^{6}$&$-2.020165(2)\times10^{6}$&$-2.020604(2)\times10^{6}$ \\
 \hline\hline
\end{longtable}
\newpage
\begin{longtable}{l c c c}
\caption{\label{g3} The long-range interaction coefficients,
 in
atomic units, for the
He($2\,^3\!P$)-He($2\,^3\!P$) system .
}\\
\hline\hline \multicolumn{1}{l}{Mass}&
\multicolumn{1}{l}{$^\infty\!\,$He($2\,^3\!P$)--$^\infty\!\,$He($2\,^3\!P$)}&
\multicolumn{1}{l}{$^4\!\,$He($2\,^3\!P$)--$^4\!\,$He($2\,^3\!P$)}& \multicolumn{1}{l}{$^3\!\,$He($2\,^3\!P$)--$^3\!\,$He($2\,^3\!P$)}\\
\hline
 $C_{5}(\Delta,0)$&$-158.044\,907\,607\,9(7)$&$-157.995\,476\,272\,6(5)$&$-157.979\,302\,728\,9(2)$\\
$ C_{5}(\Pi,+,0)$&632.179\,630\,432(3)&631.981\,905\,089(1)&631.917\,210\,917(2)\\
$C_{5}(\Pi,-,0)$&0&0&0\\
$C_5(\Sigma,0; \lambda_1)$&0&0&0\\
$C_5(\Sigma,0; \lambda_2)$&$-948.269\,445\,647(4)$&$-947.972\,857\,636(3)$&$-947.875\,816\,375(3)$\\
$C_5(\Sigma,0; \lambda_3)$&0&0&0\\
 $C_{6}(\Delta,0)$&4083.026\,998\,9(3)&4082.935\,140\,9(1)&4082.904\,962\,9(7)\\
$ C_{6}(\Pi,+,0)$&2324.937\,573\,2(4)&2325.200\,148\,8(4)&2325.285\,984\,8(3) \\
$ C_{6}(\Pi,-,0)$& 8172.742842(3)&8172.314300(1)&8172.173845(3) \\
$C_{6}(\Sigma,0; \lambda_1)$&7296.821\,336(5)&7296.447\,426(2)&7296.324\,875(5) \\
$C_{6}(\Sigma,0; \lambda_2)$&8367.590\,5(5)&8361.204\,8(2)&8359.115\,7(4) \\
$C_{6}(\Sigma,0; \lambda_3)$&1159.124\,365\,8(5)&1159.378\,065\,7(5)&1159.461\,0336(4) \\
$C_{8}(\Delta,0)$&$-30170.71(6)$&$-30168.72(8)$&$-30168.10(6)$\\
$ C_{8}(\Pi,+,0)$& 17083.8(2) &17044.4(2)&17031.5(3)\\
$ C_{8}(\Pi,-,0)$&726869.3(2)&726728.9(2)&726682.8(1)  \\
$C_{8}(\Sigma,0; \lambda_1)$&618334.9(1)&618201.6(2)&618157.8(1) \\
$C_{8}(\Sigma,0; \lambda_2)$&1835912.3(2)&1835108.6(4)&1834846(1) \\
$C_{8}(\Sigma,0; \lambda_3)$ &$-8209.2(2)$&$-8229.3(3)$&$-8236.16(8)$\\
$C_{10}(\Delta,0)$&1537493(3)&1536725(3)&1536474(3)     \\
$ C_{10}(\Pi,+,0)$&$ 3.656172(2)\times10^{7}$&$3.653559(2)\times10^{7}$&$3.652704(2)\times10^{7}$\\
$ C_{10}(\Pi,-,0)$&$8.4139100(8)\times10^{7}$&$8.410605(1)\times10^{7}$&$8.409524(2) \times10^{7}$  \\
$C_{10}(\Sigma,0; \lambda_1)$&$9.846488(2)\times10^{7}$&$9.842827(2)\times10^{7}$&$9.841629(2)\times10^{7}$ \\
$C_{10}(\Sigma,0; \lambda_2)$&$3.9464965(4)\times10^{8}$&$3.94483100(3)\times10^{8}$&$3.9442865(4)\times10^{8}$ \\
$C_{10}(\Sigma,0; \lambda_3)$&$-1.378107(3)\times10^{6}$&$-1.378199(2)\times10^{6}$&$-1.378225(5)\times10^{6}$ \\
 \hline\hline
\end{longtable}
\begin{longtable}{lll}
\caption{\label{compare} Comparison of the  present results with
the available results of Ref.~\protect\cite{Ovs82} for the $C_6$
coefficients of  He($2\,^3\!P_2$)--He($2\,^3\!P_2$) with infinite
nuclear mass. The relation between the present results (LHS) and
the results from   Ref.~\protect\cite{Ovs82} (RHS) are given in
the first column. The symbols for the dispersion coefficients on
the RHS represent values $C_6(M_A, M_B)$ for two atoms in the
total angular momentum states
 $J_A = J_B = 2$ and quantization axis along
the inter-nuclear axis.
}\\
\hline\hline
\multicolumn{1}{c}{Terms}     &
  \multicolumn{1}{c}{Present} &
  \multicolumn{1}{c}{ Ref.~\protect\cite{Ovs82}} \\
\hline
$C_{6}(\Delta,0) = C_6(2,2)$
      &  4\,083 & 4\,056    \\
$\frac 12\left[C_{6}(\Pi,+,0) + C_{6}(\Pi,-,0)\right] = 2 C_6(1,2)
- C_6(2,2)$
      & 4\,666 & 4\,645  \\
$\frac 13 \sum_{i=1}^3 C_{6}(\Sigma,0; \lambda_i) = 4C_6(0,2) +
\textstyle{\frac{4}{3}}C_6(1,1) - \textstyle{\frac{20}{3}}
C_6(1,2) + \textstyle{\frac{7}{3}} C_6(2,2) $
      & 5\,608 &   7\,026 \\
\hline\hline
\end{longtable}
\end{document}